# On spatial resolution, signal-to-noise and information capacity of linear imaging systems

T.E. Gureyev[1,2,3,4], Ya.I. Nesterets[4,3] and F. de Hoog[5]

[1]ARC Centre of Excellence in Advanced Molecular Imaging, The University of Melbourne, Parkville, VIC 3010, Australia
[2]Monash University, Clayton, VIC 3800, Australia
[3]University of New England, Armidale, NSW 2351, Australia
[4]Commonwealth Scientific and Industrial Research Organisation, Clayton, VIC 3168, Australia
[5]Commonwealth Scientific and Industrial Research Organisation, Canberra, ACT 2601, Australia

**Abstract**

A simple model for image formation in linear shift-invariant systems is considered, in which both the detected signal and the noise variance are varying slowly compared to the point-spread function of the system. It is shown that within the constraints of this model, the square of the signal-to-noise ratio is always proportional to the "volume" of the spatial resolution unit. In the case of Poisson statistics, the ratio of these two quantities divided by the incident density of the imaging particles (e.g. photons) represents a dimensionless invariant of the imaging system, which was previously termed the intrinsic imaging quality. The relationship of this invariant to the notion of information capacity of communication and imaging systems, which was previously considered by Shannon, Gabor and others, is investigated. The results are then applied to a simple generic model of quantitative imaging of weakly scattering objects, leading to an estimate of the upper limit for the amount of information about the sample that can be obtained in such experiments. It is shown that this limit depends only on the total number of imaging particles incident on the sample, the average scattering coefficient, the size of the sample and the number of spatial resolution units.

## 1. Introduction

Among the most important characteristics of many imaging, communication and measurement systems are their capacity to encode or transmit large quantities of information, resolve sufficiently fine spatial details and provide output with high signal-to-noise ratio (SNR) at lowest possible radiation doses or input energy levels [1, 2]. For mainly historical reasons (abundance of photons in typical visible light imaging applications), imaging performance characteristics related to the incident particle fluence have not been extensively studied in classical optics.

However, such characteristics, and the interplay between them, have attained additional relevance in recent years in the context of biomedical imaging, where the samples are often sensitive to the radiation doses [3], in certain astronomical methods where the detectable photon flux can be extremely low [4], as well as in some problems related to foundations of optics and quantum physics [5, 6]. In X-ray medical imaging, in particular, it is critically important to minimize the radiation dose delivered to the patient, while still being able to obtain 2D or 3D images with sufficient spatial resolution and SNR in order to detect the features of interest, such as small tumours [7, 8]. In this context, an imaging system (e.g. a CT scanner) must be able to maximize the amount of relevant information that can be extracted from the collected images, while keeping sufficiently low the number of X-ray photons impinging on the patient. The present paper addresses some mathematical properties of generic imaging systems, that are likely to be important in the context of designing medical imaging instruments, and may also have relevance to some fundamental aspects of quantum physics and information theory.

We have recently introduced a notion of intrinsic imaging quality, which incorporates both the spatial resolution and the noise propagation properties of an imaging system [6-10]. Compared to previous publications on this and related topics, Section 2 of this paper extends the notion of intrinsic imaging quality from "flat-field" (featureless) images to the situation were the signal and noise levels can vary across the images, but the rate of variation is slow compared to that of the point-spread function (PSF) of the imaging system. In Section 3, we consider the notion of information capacity of imaging systems, leading to results which are consistent with the ones previously reported on the basis of the classical approach originated by C. Shannon in the context of communication systems [11-13]. In Section 4, we apply the obtained results for the intrinsic imaging quality and information capacity of linear shift-invariant (LSI) systems to analysis of a simple model of quantitative imaging of weakly scattering samples. In particular, we derive an estimate for the maximum information about the sample that can be extracted in the relevant scattering experiments. We show that such a limit for the extractable information depends only on the total number of imaging particles used (i.e. on the radiation dose), the number of spatial resolution units and the "scattering power" of the sample, which is defined as a product of the average scattering coefficient and the linear size of the sample. The main results of the present paper are summarized in Section 5.

## 2. Noise and spatial resolution

Consider a simple model for formation of images in an LSI system which involves a stochastic flux of imaging particles incident on a position-sensitive detector, and a linear filter function describing the deterministic post-detection action of the imaging system. Here the fluence of particles (i.e. the particle density) in an image registered by the detector will typically result from interaction of particles emitted by a source with a scattering object (sample) that is being imaged, but we leave the analysis of effects of such interaction until Section 4 below.

Different realizations (instances) of images registered by the detector under the same conditions correspond to different sample functions of the stochastic detected particle fluence distribution $S_{in}(\mathbf{x})$, $\mathbf{x} \in \mathbb{R}^n$, where $n$ is the dimensionality of the image. Note that here we generalize the concept of a "detection system" to any positive integer dimension $n$, which can be convenient e.g. for description of 3D imaging. We will call the expectation value (ensemble average) $\overline{S}_{in}(\mathbf{x})$ the "detected signal" (which serves as an "input" for the post-detection signal processing system). The difference $\tilde{S}_{in}(\mathbf{x}) = S_{in}(\mathbf{x}) - \overline{S}_{in}(\mathbf{x})$ is the corresponding "noise process".

The second-order correlation properties of the particle fluence registered by the detector are described by the autocovariance function $\Gamma(\mathbf{x}, \mathbf{y}) = \overline{\tilde{S}_{in}(\mathbf{x}) \tilde{S}_{in}(\mathbf{y})}$. We assume that the autocovariance function has the following form:

$$\Gamma(\mathbf{x}, \mathbf{y}) = \frac{H(\mathbf{x} - \mathbf{y})}{H(\mathbf{0})} \sigma_{in}^2 \left( \frac{\mathbf{x} + \mathbf{y}}{2} \right), \qquad (1)$$

where the ratio $H(\mathbf{x} - \mathbf{y}) / H(\mathbf{0})$ has the meaning of the degree of correlation, and $\sigma_{in}^2(\mathbf{x}) = \Gamma(\mathbf{x}, \mathbf{x})$ is the noise variance. Note that this model is similar in form to that of the quasi-homogeneous source [2], but while the latter model represents the second-order correlation function of complex amplitudes, the present model, eq.(1), is of the fourth order with respect to the complex amplitudes. The key assumption that we make here is that the function $H(\mathbf{x})$ is narrow compared to the detected signal $\overline{S}_{in}(\mathbf{x})$ and the noise variance $\sigma_{in}^2(\mathbf{x})$, i.e. $\overline{S}_{in}(\mathbf{x})$ and $\sigma_{in}^2(\mathbf{x})$ are almost constant over the distances equal to the width of $H(\mathbf{x})$. This means that the autocovariance function $\Gamma(\mathbf{x}, \mathbf{y})$ is quasi-stationary, i.e. it can be written in the form $\Gamma_{\mathbf{x}}(\mathbf{h}) = \Gamma(\mathbf{x} + \mathbf{h}/2, \mathbf{x} - \mathbf{h}/2) = \sigma_{in}^2(\mathbf{x}) H(\mathbf{h}) / H(\mathbf{0})$ with a slowly varying "envelope" determined by $\sigma_{in}^2(\mathbf{x})$ and a rapidly varying component defined by $H(\mathbf{h})$. This allows us to introduce, by means of the Wiener-Khinchin theorem, the corresponding noise power spectrum (which is also a slowly varying function of $x$):
$W_{\mathbf{x}}(\xi) = \hat{\Gamma}_{\mathbf{x}}(\xi) = \sigma_{in}^2(\mathbf{x}) \hat{H}(\xi) / H(\mathbf{0})$, where the hat symbol denotes the Fourier transform with respect to $\mathbf{h}$.

As the noise power spectrum $W_x(\xi)$ is a non-negative and even function of $\xi$, the same is true for $\hat{H}(\xi)$. Therefore, there exists a real-valued even function $P(\mathbf{x})$, such that $\hat{H}(\xi) = |\hat{P}(\xi)|^2$, and hence $H(\mathbf{x}) = \int P(\mathbf{y})P(\mathbf{x}+\mathbf{y})d\mathbf{y}$. The function $P(\mathbf{x})$ represents the "native" point-spread function (PSF) of the imaging system, which can reflect both the correlation properties of the incident particle fluence (i.e. the properties of the source and the imaging set-up) and the PSF of the detector. We assume here that $P(\mathbf{x})$ is non-negative. Then, taking into account that $\hat{H}(\mathbf{0}) = \|H\|_1 = \|P\|_1^2$ and $H(\mathbf{0}) = \|P\|_2^2$, where $\|f\|_p = (\int |f(\mathbf{x})|^p\, d\mathbf{x})^{1/p}$ is the usual $L_p$ norm, we obtain in particular that
$W_\mathbf{x}(\mathbf{0}) = \int \Gamma_\mathbf{x}(\mathbf{h})d\mathbf{h} = \sigma_{in}^2(\mathbf{x})\hat{H}(\mathbf{0})/H(\mathbf{0}) = \sigma_{in}^2(\mathbf{x})\|P\|_1^2/\|P\|_2^2$. According to its definition and the properties of the covariance function assumed above, the native PSF $P(\mathbf{x})$ is much narrower than both the detected signal $\bar{S}_{in}(\mathbf{x})$ and the noise variance $\sigma_{in}^2(\mathbf{x})$. Note that such a behavior is "natural" for a well-designed imaging system: it ensures, in particular, that the detector PSF does not significantly smear typical incident signals. Model examples of the native PSF are represented by
$P_{Gauss}(\mathbf{x}) = (2\pi)^{-n/2}\sigma^{-n}\exp[-|\mathbf{x}|^2/(2\sigma^2)]$ and $P_{rect}(\mathbf{x}) = (2\sigma)^{-n}\chi_{[-\sigma,\sigma]^n}(\mathbf{x})$,
where $\chi_{[-\sigma,\sigma]^n}(\mathbf{x})$ is the function that is equal to one in the cubic area around the origin of coordinates with side length equal to $2\sigma$, and is equal to zero outside this area. An autocovariance function of the type described above is obtained, for example, in the case of a filtered Poisson point process [1] when the width of the filter is small compared to the typical variation length of the mean photon fluence.

We want to find out how the output of the imaging system changes after the action of an LSI transformation with a non-negative PSF (filter function) $T(\mathbf{x})$,

$$S(\mathbf{x}) = (T * S_{in})(\mathbf{x}) = \int T(\mathbf{x}-\mathbf{y})S_{in}(\mathbf{y})d\mathbf{y}. \qquad (2)$$

Let us calculate the average signal (i.e. the expectation) and the (noise) variance of the filtered process in eq.(2) under the assumption that $T(\mathbf{x})$ is varying slowly compared to $H(\mathbf{x})$ (i.e. $T(\mathbf{x})$ is almost constant over distances comparable with the correlation length of the input noise), but the width of $T(\mathbf{x})$ is much smaller than the characteristic length of variation of the detected signal $\bar{S}_{in}(\mathbf{x})$ and noise variance $\sigma_{in}^2(\mathbf{x})$. Under these assumptions, we have:

$$\overline{S}(\mathbf{x}) = \int \overline{S}_{in}(\mathbf{x}-\mathbf{y})T(\mathbf{y})d\mathbf{y} \cong \overline{S}_{in}(\mathbf{x}) \|T\|_1, \qquad (3)$$

and

$$\sigma^2(\mathbf{x}) = \iint \Gamma(\mathbf{x}-\mathbf{y}, \mathbf{x}-\mathbf{y}')T(\mathbf{y})T(\mathbf{y}')d\mathbf{y}d\mathbf{y}' \cong \sigma_{in}^2(\mathbf{x})\frac{\|H\|_1}{H(\mathbf{0})}\|T\|_2^2. \qquad (4)$$

Therefore, the following expression can be given for the output SNR:

$$SNR^2(\mathbf{x}) = \frac{\overline{S}^2(\mathbf{x})}{\sigma^2(\mathbf{x})} = SNR_{in}^2(\mathbf{x})\frac{\|P\|_2^2 \|T\|_1^2}{\|P\|_1^2 \|T\|_2^2}, \qquad (5)$$

where $SNR_{in}^2 = \overline{S}_{in}^2(\mathbf{x})/\sigma_{in}^2(\mathbf{x})$.

Having introduced the notion of SNR in the LSI system defined by eq.(3), we now turn to the notion of spatial resolution. The intrinsic spatial resolution $\Delta x$ of a system described by eq.(3) can be defined via the variance of its PSF as follows,

$$\Delta x = \left(\frac{4\pi}{n}\frac{\int |\mathbf{x}|^2 T(\mathbf{x})d\mathbf{x}}{\int T(\mathbf{x})d\mathbf{x}}\right)^{1/2} = \left(\frac{4\pi}{n}\frac{\||\cdot|^2 T\|_1}{\|T\|_1}\right)^{1/2}, \qquad (6)$$

where the factor $4\pi/n$ is introduced for normalization purposes (see Table 1), and we have assumed for simplicity that $T(\mathbf{x}) \geq 0$ is properly centred with respect to $\mathbf{x} = \mathbf{0}$, i.e. its first integral moment is equal to zero.

Following the ideas discussed in [2, 14] and elsewhere, it is also possible to define the spatial resolution of the LSI system by an alternative expression:

$$\Delta_2 x = (\|T\|_1^2 / \|T\|_2^2)^{1/n}. \qquad (7)$$

The results of calculation of $(\Delta x)^2$ and $(\Delta_2 x)^2$ for some popular PSFs shown in Table 1 demonstrate that the spatial resolution defined by eq.(7) produces similar, and in some cases more natural results, compared to the more conventional definition given by eq.(6). Note in particular that in the case of our model PSF functions $P_{Gauss}(\mathbf{x})$ and $P_{rect}(\mathbf{x})$, we get $(\Delta_2 x)_{in} = 2\sqrt{\pi}\,\sigma$ and $(\Delta_2 x)_{in} = 2\sigma$, respectively, which correlates well with the usual understanding of the "width" of these two functions.

Combining eqs. (5) and (7), we can obtain:

$$\frac{SNR^2(\mathbf{x})}{(\Delta_2 x)^n} = \frac{SNR_{in}^2(\mathbf{x})}{(\Delta_2 x)_{in}^n}. \tag{8}$$

where $(\Delta_2 x)_{in} = (\|P\|_1^2 / \|P\|_2^2)^{1/n}$ is the spatial resolution associated with the native PSF $P(\mathbf{x})$. Equation (8) means that the ratio of the square of SNR to the volume of the spatial resolution unit is invariant with respect to actions of LSI systems under the assumed conditions. In other words, an exact duality exists between the signal-to-noise and the spatial resolution of the imaging system. It can be verified that any additional convolution with another filter $V(\mathbf{x})$, which is much wider than $T(\mathbf{x})$ but much narrower than the typical variation length of the detected signal, will result in the same equation for the noise variance as eq.(5), with $T(\mathbf{x})$ replaced by $V(\mathbf{x})$. This can be proved using the fact that the two subsequent convolutions are equivalent to the convolution of the original image with the PSF $(T*V)(\mathbf{x})$, $\|T*V\|_1^2 = \|T\|_1^2 \|V\|_1^2$, and under the stated assumptions, we get also that $\|T*V\|_2^2 \cong \|T\|_1^2 \|V\|_2^2$.

If the detector area $A^n$ is fixed, the spatial resolution "area" and the total number $M_2$ of resolvable effective "pixels" (resolution units) are inversely proportional to each other: $M_2 = A^n / (\Delta_2 x)^n$. Then eq.(8) implies that the product of the number of resolvable spatial units (pixels) and the square of the SNR is also invariant under the action of LSI systems with suitable PSFs:

$$M_2 SNR^2(\mathbf{x}) = M_{2,in} SNR_{in}^2(\mathbf{x}). \tag{9}$$

Equation (9) shows that, at least under the stated assumptions, the invariant, $M_2 SNR^2(\mathbf{x})$, does not change after post-processing operations that can be described in the form of a convolution or deconvolution equation.

Note that, while the invariant defined by eq.(9) is dimensionless, the invariant defined by eq.(8) has the dimensionality of inverse length taken to the power of $n$. One way to "naturally" normalize this function and make it dimensionless is to divide it by $\bar{S}_{in}(\mathbf{x})$. Indeed, in the case of a filtered Poisson point process with the filter function $T(\mathbf{x})$ which is much narrower than the variation length of detected signal $\bar{S}_{in}(\mathbf{x})$ and the noise variance $\sigma_{in}^2(\mathbf{x})$, one has [1, Sect.11.3.9]: $\bar{S}(\mathbf{x}) = (T * \bar{S}_{in})(\mathbf{x}) \cong \|T\|_1 \bar{S}_{in}(\mathbf{x})$ and $\sigma^2(\mathbf{x}) = (T^2 * \bar{S}_{in})(\mathbf{x}) \cong \|T\|_2^2 \bar{S}_{in}(\mathbf{x})$. Then using eq.(7) we obtain:

$$\frac{SNR^2(\mathbf{x})}{(\Delta_2 x)^n \bar{S}_{in}(\mathbf{x})} = 1. \tag{10}$$

More generally, the quantity defined in eq.(8) can be expressed as $SNR_{in}^2(\mathbf{x})/(\Delta_2 x)_{in}^n = \bar{S}_{in}^2(\mathbf{x}) \|P\|_2^2 /(\sigma_{in}^2(\mathbf{x})\|P\|_1^2) = \bar{S}_{in}^2(\mathbf{x})/W_\mathbf{x}(\mathbf{0})$, and hence the latter expression gives the "normalization factor" that correctly accounts both for the strength of the detected signal and for its variation length scale. For Poisson processes one has $\bar{S}_{in}^2(\mathbf{x})/W_\mathbf{x}(\mathbf{0}) = \bar{S}_{in}(\mathbf{x})$, which coincides with the normalization factor chosen above for this case.

We have previously introduced and studied [6-10] the following dimensionless quantity, which is close in form to eq.(10) and incorporates both the noise propagation and the spatial resolution properties of LSI system:

$$Q_S^2 \equiv \frac{SNR^2(\mathbf{x})}{(\Delta x)^n \bar{S}_{in}(\mathbf{x})}. \tag{11}$$

This quantity was termed the "intrinsic imaging quality" characteristic (per single particle) of the imaging system. Note that although both $SNR$ and $\bar{S}_{in}$ can vary across the image, they have been assumed to be "quasi-stationary", i.e. approximately constant within the width of the filter function of the LSI system. It turns out that under these conditions, the imaging quality $Q_S$ is constant across the image. Indeed, substituting (10) into (11) and then taking into account the definitions of the spatial resolution from eqs.(6) and (7), we obtain

$$Q_S^2 = \frac{(\Delta_2 x)^n}{(\Delta x)^n} = \frac{\|T\|_1^{2+n/2}}{(4\pi/n)^{n/2} \||\cdot|^2 T\|_1^{n/2} \|T\|_2^2}, \tag{12}$$

i.e. the intrinsic imaging quality characteristic can be expressed purely in terms of the filter PSF alone. It is easy to verify that the functional in the right-hand side of eq.(12) is bi-invariant with respect to multiplication of the PSF or its argument by any positive constant. Therefore, $Q_S$ is independent from the height or width of the system's PSF, and depends only on its functional form (see Table 1). It was proven in [9] that this functional is always bounded from above, i.e. the inequality $Q_S^2(T) \leq 1/C_n$ holds and is exact for LSI systems with point-spread functions $T(\mathbf{x})$ having finite mathematical expectation, variance and energy, where $C_n = 2^n \Gamma(n/2) n(n+2)/(n+4)^{n/2+1}$ is the Epanechnikov constant [15, 9, 7]. The maximum, i.e. the equality $Q_S^2(T) = 1/C_n$, is achieved on Epanechnikov PSFs $T_E(\mathbf{x}) = (1 - |\mathbf{x}|^2)_+$ [15, 9], where the subscript "+" denotes that $T_E(\mathbf{x}) = 0$ at points where the expression in brackets is negative. Note that $C_1 = 6\sqrt{\pi}/125 \cong 0.95$ and $C_n \to 0$ monotonically when $n \to \infty$. Although the definition of the intrinsic quality was originally introduced for LSI systems [7], later we extended it to some non-linear systems. One such example, studied in [6], corresponded to the case of the ideal observer SNR [1] in the famous Young double-slit diffraction experiment.

We will also need an estimate for the average value of $SNR^2(\mathbf{x})$ over all spatial resolution units in the image. In the case of incident fluences with Poisson statistics, we obtain from eq.(10): $SNR^2(\mathbf{x}_m) = (\Delta x_2)^n \overline{S}_{in}(\mathbf{x}_m) = \overline{n}(\mathbf{x}_m)$, which can be identified with the average number of imaging particles registered in the "pixel" (spatial resolution unit) with the center at $\mathbf{x} = \mathbf{x}_m$. Let us sum up this quantity over all points $\mathbf{x}_m$, $m = 1, 2, ..., M_2$, corresponding to the centers of the output spatial resolution units:

$$M_2 \langle SNR^2 \rangle = \sum_{m=1}^{M_2} SNR^2(\mathbf{x}_m) = \sum_{m=1}^{M_2} (\Delta x_2)^n \overline{S}_{in}(\mathbf{x}_m) = \sum_{m=1}^{M_2} \overline{n}(\mathbf{x}_m) = N, \text{ i.e.}$$

$$\langle SNR^2 \rangle = N / M_2, \tag{13}$$

where $N$ is the average total number of incident particles per registered image and the angular brackets denote the average over all resolution units in the sample.

A useful expression for the intrinsic imaging quality characteristic can be obtained by first noting that $Q_S^2 = (\Delta_2 x)^n / (\Delta x)^n = M / M_2$, where $M = A^n / (\Delta x)^n$, and then substituting the expression $M_2 = N / \langle SNR^2 \rangle$ obtained from eq.(13), with the result given by:

$$Q_S^2 = (M / N) \langle SNR^2 \rangle. \tag{14}$$

As discussed in the next section, eq.(14) is somewhat similar in form to the expression for information capacity of imaging systems.

### 3. Information capacity of imaging systems

Consider a simple "imaging" experiment that involves $N$ incident particles ("imaging quanta", e.g. photons) and a 2D detector with $M$ pixels (where both $N$ and $M$ are non-negative integer numbers). We would like to calculate first how many distinct "images" can be produced in this situation, in other words, how many distinct sequences $(n_1, n_2, ..., n_M)$ exist, with $\sum_{m=1}^{M} n_m = N$ (all $n_m$ are non-negative integers). Two sequences $(n_1, n_2, ..., n_M)$ and $(n'_1, n'_2, ..., n'_M)$ are considered distinct if $n_m \neq n'_m$ for at least one index $m$, $1 \leq m \leq M$. This problem is equivalent to the one about the number of different ways for distributing $N$ indistinguishable (identical) balls into $M$ distinguishable bins. The corresponding total number of different combinations, that we denote by $X(M, N)$, can be found by utilizing the following recursive relationship:

$$X(M, N) = \sum_{n=0}^{N} X(M-1, n). \qquad (15)$$

Equation (15) can be understood by considering the last ($M$-th) bin: obviously, it can contain $n_M = 0$, or $1, ..., $ or $N$ balls, and in each such case, the remaining $(N - n_M)$ balls can be distributed arbitrarily between the first ($M$ - 1) bins. Comparing eq.(15) with the known theorem about the "rising sum of binomial coefficients" (see e.g. [16]), which states:

$$\sum_{n=0}^{N} \binom{M+n}{M} = \binom{M+1+N}{M+1} = \binom{M+1+N}{N}, \qquad (16)$$

we can conclude that the quantity

$$X(M, N) = \binom{M-1+N}{M-1} = \frac{(M-1+N)!}{(M-1)!N!} \qquad (17)$$

satisfies the recursive relationship of eq.(15). The correctness of "normalization" in eq.(17) is justified by considering the case of $M = 1$ bins: obviously, in this case the number of possible different placements of balls is equal to 1 (they can only all go into the one available bin). This agrees with the fact that, according to eq.(17), $X(1, N) = 1$ for any $N$.

Now let us consider the possibility that some of the particles incident on the object can be absorbed or strongly deflected by the object, etc., and as a result will not be registered by the detector. This means that some of the resultant "images" in this situation can be created with a different number $N'$ of registered quanta, where $N'$ can be any integer satisfying $0 \leq N' \leq N$. Let us calculate the number, $Y(M, N)$, of all distinct images that can be formed this way. Obviously,

$$Y(M,N) = \sum_{N'=0}^{N} X(M,N'). \qquad (18)$$

Knowing the explicit form of $X(M,N)$ given by eq.(17), we can use eq.(16) again to evaluate the sum on the right-hand side of eq.(18), with the result:

$$Y(M,N) = \binom{M+N}{M} = \binom{M+N}{N} = \frac{(M+N)!}{M!\,N!}. \qquad (19)$$

Note, in particular, that in the simplest case of one bin, $M = 1$, eq.(19) gives $Y(1,N) = N+1$. This is the correct result, as the only bin available in this case can contain any number of balls between 0 and $N$.

A simpler approximation for the expression $Y(M,N)$ in eq.(19) can be derived for large values of $M$ and $N$. According to the refined version of Stirling's formula for the factorial [17]:

$$n! = \sqrt{2\pi}\, n^{n+1/2} e^{-n} e^{w_n}, \; n = 1, 2, ..., \text{ where } \frac{1}{12n+1} < w_n < \frac{1}{12n}. \qquad (20)$$

Let us introduce a notion of "information capacity" $D(M,N) = \log_2 Y(M,N)$ by analogy with the original information capacity definition introduced by Shannon [11] for communication systems and later extended to imaging systems [12-13]. We will also consider information capacity per single particle:
$D_S(M,N) = D(M,N)/N = (1/N)\log_2 Y(M,N)$.

From eqs.(19)-(20) we obtain:

$$\begin{aligned} D(M,N) &= (M+N+1/2)\log_2(M+N) - (M+1/2)\log_2 M \\ &\quad -(N+1/2)\log_2 N + O(1), \end{aligned} \qquad (21)$$

where $O(1)$ does not exceed a constant of the order of 1 for any non-negative integer $M$ and $N$. It may be interesting to note that the expressions for $Y(M,N)$ and $D(M,N)$ are symmetric with respect to $M$ and $N$, even though by definition all the imaging particles are considered identical (indistinguishable) and all the detector pixels are uniquely identifiable (distinguishable).

Usually in practice the number of imaging particles (e.g. photons) in an image is much larger than the number of detector pixels, which corresponds to the case $N \gg M$ in the above equations. In that case eq.(21) leads to the following expressions

$$D(M,N) = M[\log_2(N/M) + O(1)], \qquad (22)$$

$$D_S(M,N) = (M/N)[\log_2(N/M) + O(1)], \text{ when } N/M \to \infty. \qquad (23)$$

Let us now postulate that two images are distinguishable if and only if the difference between their signal values in at least one pixel exceeds the sum of standard deviations of noise in that pixel in the two images (this can be easily modified to include a distinguishability factor $SNR_{min}$ which can e.g. be equal to 5, if the Rose criterion is applied). In the "noise-free" case considered above, the distinguishability was equivalent to any difference in detected particle numbers in the two images by at least one particle in at least one pixel, which formally corresponds e.g. to the standard deviation of noise being equal to 1/2 or less in all pixels of both images.

Consider the case where the detector is not perfect and contains a fixed amount of additive noise in each pixel. Assuming that the level of noise is $\sigma$, two signals, corresponding to $n_1$ and $n_2$ registered particles in a pixel, respectively, can be considered distinguishable e.g. if $|n_1 - n_2| \geq 2\sigma$. This is equivalent to $|n_1/(2\sigma) - n_2/(2\sigma)| \geq 1$. In other words, the number of distinguishable images can now be calculated in terms of "bunches" of $2\sigma$ particles, instead of the individual particles (note that the fractions of the bunches, that correspond to signal increments below the level of distinguishability, may lead to a reduction of the total number of full bunches). Therefore, the result obtained above in eqs. (22)-(23) should remain valid in the case of additive detector noise, with the replacement of $N$ by $N/(2\sigma)$. Note also that

$\log_2(N/(2\sigma M)) = \log_2(N/M) - \log_2(2\sigma) = \log_2(N/M) + O(1)$, and hence the form of eqs.(22-23) can remain unchanged in this case.

Finally consider the case of a photon-counting detector, which corresponds to the Poisson statistics with the expectation and variance both equal to the mean number $\bar{n}$ of photons per pixel, $\sigma = \sqrt{\bar{n}}$. Let us define two signal levels as distinguishable if they differ at least by the sum of their standard deviations. If the number of photons in a pixel is equal or smaller than $\bar{n}$, then there are $\sqrt{\bar{n}}$ distinguishable possible signal levels. For example, for $\bar{n} = 25$, the $\sqrt{\bar{n}} = 5$ distinguishable levels (bands) correspond to 25±5, 16±4, 9±3, 4±2, and 1±1 photons. Therefore, the result stated in eqs.(22)-(23) should remain valid in the case of Poisson statistics, with the replacement of $N$ by $\sqrt{N}$. Note also that
$M \log_2(\sqrt{N}/M) + O(M) = 0.5 M \log_2(N/M) + O(M \log_2 M)$.

Although the above considerations regarding the extension of eqs.(22)-(23) to the case of noisy images are not rigorous, it turns out that they lead to correct formulae for the information capacity in the case of Poisson (shot) and Gaussian (additive) noise. A rigorous derivation for these cases is given in the Appendix where more accurate formulae are derived from the results in [18-20]. Specifically, an estimate for the information capacity for the case of Poisson statistics when $N/M$ is large, is given by

$$D_P(M,N) = 0.5M[\log_2(N/M) + o(1)], \tag{24}$$

from which it follows that

$$D_{P,S}(M,N) = 0.5(M/N)[\log_2(N/M) + o(1)], \tag{25}$$

where the term $o(1)$ tends to zero in the limit as $N/M \to \infty$.

Identifying the individual "pixels" with the isotropic spatial resolution units having the side length given by eq.(7) and substituting eq.(13) into eq.(24), we obtain that in the case of Poisson noise:

$$D_P(M_2,N) = 0.5M_2[\log_2\langle SNR^2\rangle + o(1)]. \tag{26}$$

where the term $o(1)$ tends to zero in the limit as $N/M_2 \to \infty$. The corresponding equation in the case of Gaussian additive noise is:

$$D_G(M_2,N) = 0.5M_2[\log_2(\langle SNR^2\rangle + 1)]. \tag{27}$$

Previously, expressions for the information capacity of imaging systems with additive noise in a form similar to eq.(27) have been obtained in [12, 13].

### 4. Effect of the imaged sample

The above analysis has been carried out without a reference to an imaged sample. In the present section we consider one simple generic model which takes into account the interaction of the incident particle fluence with a sample and its consequences for the information about the sample that can be obtained in an imaging experiment.

Consider an experiment involving interaction of a flux of incident imaging particles with a weakly scattering sample, after which the scattered particles are registered by a position-sensitive detector. We assume for simplicity that the particle fluence over the "incident surface" of the sample can be described by its spatially uniform expectation value $\bar{S}_0^{(n-1)}$. Note that the latter quantity is expressed in the number of particles per unit "area", i.e. it has the dimensionality of inverse length in the power of $n - 1$, while the detected signal, $\bar{S}_{in}(\mathbf{x})$, introduced in Section 2 was expressed in the number of particles per unit "volume", i.e. it has the dimensionality of inverse length in the power of $n$. We assume here that the detected scattered fluence $S_{in}(\mathbf{x})$ has a Poisson distribution with the mean $\bar{S}_{in}(\mathbf{x}) = \bar{S}_0^{(n-1)}\gamma(\mathbf{x})$. This corresponds to the case of so-called "dark-field imaging", when the primary unscattered beam does not reach the detector. Modification to the case of bright-field imaging is quite straightforward and the corresponding result is given below. Here $\gamma(\mathbf{x})$ is a deterministic non-negative function ("scattering coefficient") that describes the interaction of the incident fluence with the sample. This interaction is assumed to be linear with respect to the interaction length, i.e. $\gamma(\mathbf{x})$ has the dimensionality of inverse length. As a model for $\gamma(\mathbf{x})$ one may consider, for example, the linear attenuation, $\mu(\mathbf{x}; \lambda) = (4\pi / \lambda)\beta(\mathbf{x}; \lambda)$, or refraction, $(2\pi / \lambda)\delta(\mathbf{x}; \lambda)$, coefficients in X-ray imaging, where $n = 1 - \delta + i\beta$ is the complex refractive index of the sample and $\lambda$ is the X-ray wavelength (we omit the argument $\lambda$ in the notation for $\gamma(\mathbf{x})$ for brevity). It can sometimes be helpful to express $\gamma(\mathbf{x}) = \sigma\rho(\mathbf{x})$, where $\sigma$ is the scattering cross-section and $\rho(\mathbf{x})$ is the number density of scattering centers (usually, atoms) in the sample. The interaction is also assumed to be weak, so that $\gamma(\mathbf{x})A \ll 1$, where $A$ is the linear size of the sample. In particular, this allows us to ignore the weakening of the incident flux in the course of its propagation through the sample. Finally, the interaction has a "single-scattering" nature, i.e. each incident particle interacts at most with a single scattering center the sample, as is usually assumed in the first Born approximation or in the so-called "kinematical approximation" in X-ray and electron diffraction [21]. This model is somewhat similar to the one used in diffraction tomography [22], where the first Born approximation is invoked for the description of interaction of the radiation with the sample. However, unlike diffraction tomography, our simple model ignores the effect of propagation of the scattered radiation to the detector, assuming only that there is a one-to-one correspondence between the resolution volumes in the sample and the effective "detector pixels". The totality of detector pixels may possibly be aggregated over multiple 2D exposures of the sample. For example, in conventional or diffraction CT imaging, such one-to-one correspondence may be achieved after a reconstruction procedure, e.g. after the application of the inverse Radon transform to the data collected in real detector pixels at different rotational positions of the sample. In the case of weakly absorbing or weakly scattering samples, such reconstruction procedure is usually linear and can be expressed by a convolution with the corresponding filter function [23], and hence these cases fit well into the LSI approach considered above.

Consider now a post-detection filtering of the Poisson process with a narrow filter function $T(\mathbf{x})$ as in Section 2 above (i.e. the filter function is required to be narrow with respect to the characteristic variation length of the scattering properties across the sample). We then obtain:
$\overline{S}(\mathbf{x}) = (T * \overline{S}_{in})(\mathbf{x}) \cong \|T\|_1 \overline{S}_0^{(n-1)} \gamma(\mathbf{x})$ and
$\sigma^2(\mathbf{x}) = (T^2 * \overline{S}_{in})(\mathbf{x}) \cong \|T\|_2^2 \overline{S}_0^{(n-1)} \gamma(\mathbf{x})$. Accordingly,

$$SNR^2(\mathbf{x}) = \frac{\|T\|_1^2 \overline{S}_0^{(n-1)} \gamma(\mathbf{x})}{\|T\|_2^2} = \overline{S}_0^{(n-1)} \gamma(\mathbf{x}) (\Delta_2 x)^n. \tag{28}$$

We can now use the results from the previous sections to characterize the information capacity of an LSI imaging system with a weakly scattering object. In the present context, this can be understood as a capacity of the system to distinguish between different imaged objects.

Using the expression for $SNR^2(\mathbf{x})$ from eq.(28) and arguing similarly to the derivation of eq.(13) above, i.e. averaging over all spatial resolution units into which the sample is split, we can obtain

$$\left\langle SNR^2 \right\rangle = \overline{\gamma} A \overline{S}_0^{(n-1)} A^{n-1} \frac{(\Delta_2 x)^n}{A^n} = \frac{\Omega N}{M_2}. \tag{29}$$

where $\overline{\gamma} = M_2^{-1} \sum_{m=1}^{M_2} \gamma(\mathbf{x}_m)$ is the average value of the linear scattering coefficient in the sample, and $\Omega = \overline{\gamma} A = (\sigma / A^{n-1}) N_{sc}$ can be termed the scattering power of the sample, with $N_{sc}$ denoting the total number of scattering centers in the sample and $\sigma / A^{n-1}$ being the relative scattering cross-section. It is easy to verify that in the case of bright-field imaging one obtains

$$\left\langle SNR_{b.f.}^2 \right\rangle = \Omega^{(2)} N / M_2, \text{ with } \Omega^{(2)} = \overline{\gamma^2} A^2. \tag{30}$$

Let us consider for comparison a more specific example of conventional computed tomography (CT) imaging. It was shown in [24] that in the case of filtered back-projection reconstruction (FBP) with the nearest-neighbor interpolation, one has:

$$<SNR^2> = \frac{12}{\pi^2} \frac{\overline{\mu^2} M_a \overline{S}_0^{(2)} A^4}{M_p^2} = \left(\frac{\pi}{2}\right)^{4/3} \frac{12}{\pi^2} \frac{\Omega^{(2)} N}{M_2^{4/3}}, \tag{31}$$

where $M_a$, $M_p$ and $M_2 = M_a M_p$ are, respectively, the number of projections, the number of pixels in each projection and the total number of resolution units utilized in the whole CT scan. Here we used the well-known optimal relationship, $M_a = (\pi/2) M_p^{1/2}$, [25] (which leads to isotropic spatial resolution in the reconstructed slices), for derivation of the final result in eq.(31). Note that equation (31) correctly reflects the well-known fact [3] that in order to keep the average SNR in the reconstructed CT images constant, the number $N$ of incident particles (and hence the radiation dose to the sample) has to be changing as the inverse 4th power of the spatial resolution (note that $(\Delta_2 x) \sim 1/M_2^{1/3}$). The cause of the difference between this and the corresponding result for the bright-field scattering model, eq.(30), which has a different power of $M_2$ in the denominator, is in the mathematical properties of the inverse Radon transform. The latter is known to amplify the noise in the input signal via the application of the ramp filter in the course of FBP CT reconstruction (this fact is also reflected in the growth of the eigenvalues of the inverse Radon transform with the radial index) [25].

Using eq.(29) in combination with eq.(26) for the information capacity, we obtain:

$$D_P(M_2, N) = 0.5 M_2 [\log_2(\Omega N / M_2) + o(1)]. \qquad (32)$$

When the spatial resolution is fixed, the dependence of this information capacity on the number of particles $N$ is straightforward, $D_P(M_2, N) = 0.5 M_2 \log_2 N + const$. When the number of incident particles is fixed, then the information capacity is bounded from above by a linear function of the number of spatial resolution units, $D_P(M_2, N) \leq M_2 O(1)$.

Substituting the expression for the $SNR^2$ from eq.(28), together with the expression $S_{in}(\mathbf{x}) = \bar{S}_0^{(n-1)} \gamma(\mathbf{x})$ for the incident fluence, into the definition of $Q_S^2$ in eq.(11), we find that the imaging quality here has exactly the same form as in the case with no sample considered in Section 2. In particular, eq.(12) and the estimate $Q_S^2 \leq 1/C_n$ remain unchanged. This happens because, according to eq.(12), the imaging quality can be expressed in terms of the two definitions of the system's spatial resolution alone, and these resolutions do not depend on the effect of the scattering power of the sample under the conditions of the simple model considered here.

## 5. Conclusions

We have considered the relationship between the spatial resolution, signal-to-noise ratio and information capacity of linear shift-invariant imaging systems. Convenient imaging performance characteristics of such systems include the imaging quality $Q_S$ and the information capacity per single imaging particle (e.g. photon), $D_S$. From the results of this paper it is clear that, apart from relatively insignificant factors, the information capacity per single particle and the intrinsic imaging quality have similar mathematical expressions, namely $D_S(M,N) \sim (M/N)\log_2 <SNR^2>$ and $Q_S(M,N) \sim (M/N)<SNR^2>$, respectively. However, the behavior of these two quantities under linear transformations of the imaging system is different due to the presence of the logarithm in the first expression. It is usually possible to increase the signal-to-noise (SNR) ratio at the expense of spatial resolution (or equivalently, the total number of spatial resolution units, $M$) and vice versa using operations like image filtering, denoising or deconvolution. While the intrinsic imaging quality remains essentially invariant under such (linear) transformations, the information capacity can change significantly: for example, a system with a single spatial resolution unit has quite low information capacity per particle, $\sim (\log_2 N)/N$. On the other hand, when the number of particles is fixed, the information capacity per particle can grow without a limit when the number of resolution units grows: consider the fact that an unlimited number of new images can be created, in principle, by directing a photon or a fixed number of photons into new pixels (one pixel at a time).

In imaging problems that involve quantitative analysis of samples on the basis of transmitted or, more generally, scattered radiation, the notion of "useful signal" is modified. We have studied one example of this type of problems in Section 4 of the paper. We have found that the information capacity is proportional to the logarithm of the product of the total number of imaging particles and the scattering power of the sample, the latter depending only on the average scattering coefficient and the size of the sample (or alternatively, on the total number of scattering centers, e.g. atoms, in the sample, and the average relative scattering cross-section of each such center). On the other hand, the imaging quality per single particle is not affected by the scattering properties of the sample, and remains the same as in the sample-free case. It means that, in accordance with its name and the intended role, the intrinsic imaging quality characteristic depends only on the properties of the imaging system itself, and does not depend on the imaged sample.

**Acknowledgement**

T.E.G. acknowledges helpful discussions with Drs. H. Quiney, D. Paganin, A. Kozlov, S. Soha, and A. Martin of the School of Physics, The University of Melbourne.

Table 1. Spatial resolution associated with different PSFs. All PSFs are normalized such that $\|T(\mathbf{x};\sigma)\|_1 = 1$.

| PSF | $T(\mathbf{x};\sigma)$ | $\hat{T}(\mathbf{u};\sigma)$ | $(\Delta x)^2$ | $(\Delta_2 x)^2$ | $Q_S^{4/n}$ |
|---|---|---|---|---|---|
| Sech | $\dfrac{\sigma^{-1}}{e^{\pi x/2\sigma}+e^{-\pi x/2\sigma}}$ | $\dfrac{2}{e^{2\pi\sigma u}+e^{-2\pi\sigma u}}$ | $4\pi\sigma^2$ | $\pi^2\sigma^2$ | $\dfrac{\pi}{4}$ |
| Lorentz | $\dfrac{\pi^{-1}\sigma}{\sigma^2+x^2}$ | $e^{-2\pi\sigma|u|}$ | $\infty$ | $4\pi^2\sigma^2$ | 0 |
| Exp | $(\pi/\sigma)e^{-2\pi|x|/\sigma}$ | $\dfrac{1}{1+\sigma^2 u^2}$ | $\dfrac{2}{\pi}\sigma^2$ | $\dfrac{4}{\pi^2}\sigma^2$ | $\dfrac{2}{\pi}$ |
| $n$-dim Gauss | $\dfrac{1}{(2\pi)^{n/2}\sigma^n}e^{-\frac{|\mathbf{x}|^2}{2\sigma^2}}$ | $e^{-2\pi^2\sigma^2|\mathbf{u}|^2}$ | $4\pi\sigma^2$ | $4\pi\sigma^2$ | 1 |
| $n$-dim Rect | $\dfrac{1}{(2\sigma)^n}\chi_{[-\sigma,\sigma]^n}(\mathbf{x})$ | $\displaystyle\prod_{k=1}^{n}\dfrac{\sin(2\pi\sigma u_k)}{2\pi\sigma u_k}$ | $\dfrac{4\pi}{3}\sigma^2$ | $4\sigma^2$ | $\dfrac{3}{\pi}$ |
| Epanech- nikov *) | $\dfrac{A_n}{\pi^{n/2}\sigma^n}\left(1-\dfrac{|\mathbf{x}|^2}{\sigma^2}\right)_+$ | $A_n\dfrac{J_{n/2+1}(2\pi\sigma|\mathbf{u}|)}{(\pi\sigma|\mathbf{u}|)^{n/2+1}}$ | $\dfrac{4\pi}{n+4}\sigma^2$ | $\dfrac{4\pi\sigma^2}{(n+4)C_n^{2/n}}$ | $\dfrac{1}{C_n^{2/n}}$ |

*) In this row: $A_n = (n/2+1)\Gamma(n/2+1)$,
$C_n = 2^n \Gamma(n/2) n(n+2)/(n+4)^{n/2+1}$. Note that $1/C_n^{2/n} > 1$.

**Appendix A**
**Information capacity of a simple imaging system in the cases of additive and Poisson detection statistics**

Following the ideas by Shannon [11, 18] and others [12, 13], we consider the maximum number of distinct images ("messages"), that can be represented (encoded) by a system which has a given number of "pixels" (spatial resolution units), with an average level of signal-to-noise ratio squared over the image pixels not exceeding a certain value. Note that while in previous publications [11-13] the noise was assumed to be additive, here we consider the case of Poisson noise. Setting an upper limit for the average squared SNR in image pixels is then equivalent to limiting the average number of photons per image pixel. As the number of pixels here is considered constant, this in turn is equivalent to setting an upper limit on the total number of particles or on the energy used to form each image.

We begin by considering a communication system with inputs $\lambda_1, \lambda_2, \lambda_3, \cdots$ and corresponding outputs $y_1, y_2, y_3, \cdots$ and closely follow the development given in Shannon [18]. Further detail may be found in the excellent text on information theory by Cover and Thomas [19]. For a system without memory, Shannon [18] modelled the system by assuming that the inputs $\lambda_1, \lambda_2, \lambda_3, \cdots$ are independent realizations of a random variable $\Lambda$, say. The outputs $y_1, y_2, y_3, \cdots$ are then independent realizations of a random variable, $Y$ say. In his analysis of a noiseless system, Shannon showed that, for large $M$, there are approximately $2^{MH(\Lambda)}$ distinguishable sequences of the form $\lambda_1, \lambda_2, \cdots, \lambda_M$ where $H(\Lambda)$ is the entropy of $\Lambda$. It therefore follows that the average information content of a realization of $\Lambda$ is $H(\Lambda)$. As an example, consider the case where there are $r$ distinct values for $\lambda$ that are equally probable. Then, if $M$ is large, there are approximately $M^r$ distinct sequences of $\lambda_1, \lambda_2, \cdots, \lambda_M$, each with probability $M^{-r}$. Note that when $r = \bar{n} = N/M$, this is consistent with eq.(24).

When there is a one to one correspondence between the output of a channel and its input, the channel capacity $C$ is given by,

$$C = \max_{\Lambda} H(\Lambda), \qquad (A1)$$

where the maximum is taken over all random variables $\Lambda$ and may be subject to constraints such as a bound on the variance. However, in practice a communication channel will be subject to noise and this will affect its performance in a negative way. Specifically, the average information content from a single output is the mutual information $I(\Lambda;Y)$ which is defined by

$$I(\Lambda;Y) = H(\Lambda) - H(\Lambda|Y), \tag{A2}$$

where $H(\Lambda|Y)$ is the conditional entropy of $\Lambda$ given $Y$. The channel capacity is now defined by

$$C = \max_{\Lambda} I(\Lambda|Y), \tag{A3}$$

where, as previously, the maximum is taken over all random variables $\Lambda$ and may be subject to constraints.

We have noted previously, that when $M$ is large, the number of distinguishable sequences input sequences of the form $\lambda_1, \lambda_2, \cdots, \lambda_M$ is approximately $2^{MH(\Lambda)}$ and it is only these sequences that have information content. Each of these distinguishable sequences is received as an output sequence $y_1, y_2, \cdots, y_M$ and, in the process, may become indistinguishable. Shannon showed that about 1 in $2^{MH(\Lambda|Y)}$ distinguishable input sequences results in a distinguishable output sequence. That is, there are approximately $2^{MI(\Lambda,Y)}$ distinguishable output sequences that correspond to distinguishable input sequences.

We now, relate the models for communication systems to images. As previously, we have $M$ pixels and a total of $N$ photons. For each pixel, we have a target number of photons $\lambda_i$, $i = 1, \cdots, M$ which are independent realization of a random variable $\Lambda$ and there are associated outputs $y_i$, $i = 1, \cdots, M$, which are the number of photons measured in each pixel. As the mean number of photons per pixel is $\bar{n} := N/M$, we assume that

$$\mathbb{E}(\Lambda) = \bar{n}. \tag{A4}$$

It now remains to specify a model of the outputs, and here we shall consider two cases, namely:
- $y_i$ is an independent realization of a Poisson random variable with intensity $\lambda_i$
- $y_i = \lambda_i + z_i$ where $z_i$ is an independent a realization of a mean zero Gaussian random variable $Z$ that is independent of $\Lambda$.

The case of a Poisson distribution is important in a number of applications, including the capacity of a discrete time Poisson channel, and has received considerable attention in the literature. Although an explicit analytical expression for the channel capacity is not known, a number of upper and lower bounds are derived in Lapidoth and Moser [20, eqns 21 and 22]. When the mean number of photons per pixel $\bar{n}$ is large, these yield

$$C_{\text{Poisson}} = \tfrac{1}{2}\log_2(\bar{n}) + o(1) \tag{A5}$$

where the $o(1)$ term becomes zero in the limit as $\bar{n} \to \infty$. This is consistent with the results in Section 3.

The channel capacity for a communications channel with independent additive Gaussian noise, is well known [11]. In this case, we have $Y = \Lambda + Z$ where $\Lambda$ is a random variable representing the signal and $Z$ is an independent mean zero Gaussian random variable that models the noise. The channel capacity is given by

$$C_{Gauss} = \tfrac{1}{2}\log_2\left(\frac{\mathbb{E}(Y^2)}{\mathbb{E}(Z^2)}\right) = \tfrac{1}{2}\log_2\left(\frac{\mathbb{E}(\Lambda^2) + \mathbb{E}(Z^2)}{\mathbb{E}(Z^2)}\right). \tag{A6}$$

In order to compare this to the Poisson case, we take the same second order moments, namely $\mathbb{E}(Y^2) = \bar{n} + \bar{n}^2$, and $\mathbb{E}(Z^2) = \bar{n}$. This yields

$$C_{Gauss} = \tfrac{1}{2}\log_2(\bar{n}+1) = \tfrac{1}{2}\log_2(\bar{n}) + o(1), \tag{A7}$$

and is asymptotically the same as the Poisson case when the average number of photons per pixel is large.